\documentclass[preprints,article,accept,moreauthors,pdftex]{mdpi} 

\firstpage{1}
\pubvolume{xx}
\issuenum{1}
\articlenumber{5}
\pubyear{2020}
\copyrightyear{2020}
\history{Received: 29 February 2020; Accepted: 27 April 2020; Published:  5 May 2020}
\updates{yes} % If there is an update available, un-comment this line

\Title{Cosmological Finsler Spacetimes} % Attention AE/ME. The following layout issues have not been checked by the English Editing Department and must be carefully verified by the AE/Layout Department: All callout issues, bold usage of callouts, and references to callouts in the text. Correct callout usage in figures. Figure and Table layout issues. Footnote formatting and Glossaries have not been checked. En dash usage for negative values, en dash usage to indicate relationships, en dash usage to indicate bonds (especially in chemistry). The English Editing Department is not responsible for correct italic usage for genes, proteins and technical terminology. This responsibility belongs to the authors. The following are also not checked: spacing between numbers and units of measurement, ratios, en dashes for ranges, date and time formats, punctuation in equation lines, and less than/more than spacing (< >). Finally, capitalization and layout of titles/headings must be properly checked as well as ensuring 'Eq.' and 'Fig.' are properly spelled out, as these are layout issues.

\Author{Manuel Hohmann $^{1,\dagger}$, Christian Pfeifer $^{1,\dagger}{}$ and Nicoleta Voicu $^{2,}$*$^{,\dagger}$} % Please carefully check the accuracy of names and affiliations. Changes will not be possible after proofreading.

\AuthorNames{Manuel Hohmann, Christian Pfeifer and Nicoleta Voicu}

\address{%
$^{1}$ \quad Laboratory of Theoretical Physics, Institute of Physics, University of Tartu, W. Ostwaldi 1, 50411~Tartu,~Estonia; manuel.hohmann@ut.ee (M.H.); christian.pfeifer@ut.ee (C.P.)\\
$^{2}$ \quad Faculty of Mathematics and Computer Science, Transilvania University, Iuliu Maniu Str. 50, 500091~Brasov,~Romania
}

\corres{Correspondence: nico.voicu@unitbv.ro}

\firstnote{These authors contributed equally to this work.}
\abstract{Applying the cosmological principle to Finsler spacetimes, we identify the Lie Algebra of symmetry generators of spatially homogeneous and isotropic Finsler geometries, thus generalising Friedmann-Lema\^{i}tre-Robertson-Walker geometry. In particular, we find the most general spatially homogeneous and isotropic Berwald spacetimes, which are Finsler spacetimes that can be regarded as closest to pseudo-Riemannian geometry. They are defined by a Finsler Lagrangian built from a zero-homogeneous function on the tangent bundle, which encodes the velocity dependence of the Finsler Lagrangian in a very specific way. The obtained cosmological Berwald geometries are candidates for the description of the geometry of the universe, when they are obtained as solutions from a Finsler gravity equation.}

\keyword{Finsler geometry; Berwald space; Berwald spacetime; cosmology; cosmological principle on Finsler spacetimes}

\newcommand{\change}[1]{{\color{blue}#1}}

\begin{document}
	
\section{Introduction}\label{sec:intro}
To describe the evolution of the whole universe in cosmology, one applies the cosmological principle (CP), which states that there exists no preferred spatial position and no preferred spatial direction on large scales. Applying this principle to general relativity leads to the spatially homogeneous and isotropic Friedmann-Lema\^{i}tre-Robertson-Walker (FLRW) metric as the unique ansatz for the geometry of spacetime. It contains two free functions which depend only on time, the~lapse function and the scale factor. The~lapse function can be normalized to unity, by~a suitable choice of the time coordinate. The~scale factor remains the only free function to be determined as solution of the Einstein equations, sourced by a perfect fluid energy-momentum tensor. On~the basis of this mathematical model for the universe, one has to conclude that only $\sim$5\% of the Universe consists of standard model baryonic matter, while the rest is composed of what is nowadays called dark energy~\cite{Peebles:2002gy,Copeland:2006wr} and dark matter~\cite{Bergstrom:2000pn}. The~standard approach to cosmology is excellently summarized for example in reference~\cite{Weinberg}.

A promising approach for a geometric explanation of the dark matter and dark energy phenomenology is to use Finsler spacetime geometry for the description of the gravitational interaction, instead of  pseudo-Riemannian geometry~\cite{Bogoslovsky1994,Gibbons:2007iu,Kostelecky:2011qz,Pfeifer:2019wus,Lammerzahl:2018lhw}. In~particular, it has recently been suggested that Finsler geometry provides the correct mathematical framework~\cite{Hohmann:2019sni} and extension of the Einstein equations~\cite{Hohmann:2018rpp} for the accurate determination of the gravitational field distribution of a kinetic~gas.

In this article we apply the cosmological principle to Finsler spacetime geometry. Starting from a symmetry group which acts transitively on spatial equal time surfaces and which contains a local isotropy group acting transitively on spatial directions at each point, we find that a cosmological homogeneous and isotropic Finsler geometry is defined by a Finsler Lagrangian with a very specific dependence on the tangent bundle coordinates. Yet, since the Finsler Lagrangian is a $2$-homogeneous function in its dependence on the directional variable of the tangent bundle, the~demand of cosmological symmetry leaves large classes of allowed Finsler Lagrangians; the symmetry demand does not provide a strong limitation in this regard. Specific choices of Finsler geometries have been investigated in their capability to explain aspects of the cosmological dark matter and dark energy phenomenology~\cite{Papagiannopoulos:2017whb,Li:2015sja,Hohmann:2016pyt,Minas:2019urp,Ikeda:2019ckp,Hasse:2019zqi,Triantafyllopoulos:2018bli}.

A class of Finsler spacetime geometries, which can be regarded as closest to pseudo-Riemannian geometry, are the so-called Berwald spacetimes~\cite{Berwald1926,Fuster:2018djw,Szilasi2011,Gallego2017}. They are characterized by the fact that, in~any local chart, the~Chern-Rund connection coefficients on the tangent bundle only depend on the points of the base manifold, or, equivalently, the~geodesic spray coefficients are quadratic in their dependence on the directional variables of the tangent bundle. In~other words, on~Berwald spacetimes, the~Chern-Rund connection gives rise to an affine connection on the spacetime manifold. In~general, this connection is not the Levi-Civita connection of any pseudo-Riemannian metric, but~instead, it can be regarded as a non metric-compatible metric-affine connection without torsion~\cite{Fuster2020}.

We derive the most general cosmologically (spatially homogeneous and isotropic) Berwald spacetime geometry. It serves as simplest Finslerian candidate for the description of the geometry of the universe. The~obtained Finsler Lagrangian contains one free function, which encodes the velocity dependence of the Finsler Lagrangian in a very specific way. The~Berwald geometry we obtain is the minimal Finsler geometric extension of pseudo-Riemannian FLRW~geometry.

We present our results in the following way. In~Section~\ref{sec:fins}, we introduce the definition of Finsler spacetimes and the mathematical language needed to discuss the cosmological principle and cosmologically symmetric Finsler spacetimes. We apply the cosmological principle to Finsler spacetime geometry and identify the symmetry generating vector fields in Section~\ref{ssec:cosmo}. In~Section~\ref{sec:BerwCond} we recall how to identify Berwald spacetimes and derive the most general cosmologically symmetric Berwald spacetime Finsler Lagrangian. Finally, we conclude in Section~\ref{sec:conc}.

%%%%%%%%%%%%%%%%%%%%%%%%%%%%%%%%%%%%%%%%%%%%%%%%%%%%%%%%%%%%%%%%%%%%%%%%%%%%%%%%%%%%%%%
\section{Berwald Finsler Spacetime~Geometry}\label{sec:fins}
Throughout this article, we consider the tangent bundle $TM$ of a $4$-dimensional connected and oriented manifold $M$, equipped with manifold induced local coordinates, as~follows. A~point $(x, \dot x)\in TM$ will have local coordinates of the form $(x^a,\dot x^a)$, where $x^a$ are the local coordinates of the point $x \in M$ and $\dot x = \dot x^a \partial_a \in T_xM$ is the decomposition of the vector $\dot x \in T_xM$ in the natural basis.  If~there is no risk of confusion, we will sometimes suppress the indices of the coordinates. The~symbol $\pi$ denotes the canonical projection of the tangent bundle. The~local coordinate bases of the tangent and cotangent spaces, $T_{(x,\dot x)}TM$ and $T^*_{(x,\dot x)}TM$, of~the tangent bundle  are $\{\partial_a = \frac{\partial}{\partial x^a},\dot \partial_a = \frac{\partial}{\partial \dot x^a}\}$ and $\{dx^a, d\dot x^a\}$. 

A conic subbundle of $TM$ is a non-empty open submanifold $\mathcal{Q}\subset TM\backslash \{0\}$, with~the following~properties:
\begin{itemize}[leftmargin=2.3em,labelsep=5.8mm]
	\item $\pi _{TM}(\mathcal{Q})=M$;
	\item \textit{conic property:} if $(x,\dot{x})\in \mathcal{Q}$, then, for~any $\lambda >0:$ $(x,\lambda \dot{x})\in \mathcal{Q}$.
\end{itemize}

By a Finsler spacetime we will understand in the following a pair $(M,L)$, where $M$ is a smooth $n$-dimensional manifold and the Finsler Lagrangian $L:\mathcal{A} \to \mathbb{R}$ is a smooth function on a conic subbundle $\mathcal{A}\subset TM$, such~that:
\begin{itemize}[leftmargin=2.3em,labelsep=5.8mm]
	\item $L$ is positively homogeneous of degree two with respect to $\dot x$: $L(x,\lambda \dot x) = \lambda^2 L(x,\dot x)$ for all $\lambda \in \mathbb{R}^+$;
	\item on $\mathcal{A}$, the~vertical Hessian of $L$, called $L$-metric, is non-degenerate,
\begin{align}\label{g_def}
	g^L_{ab}=\dfrac{1}{2}\dfrac{\partial ^{2}L}{\partial \dot{x}^{a}\partial \dot{x}^{b}};
	\end{align}
	\item there exists a conic subset $\mathcal{T}\subset \mathcal{A}$ such that on $\mathcal{T}$, $L>0$, $g^L$ has Lorentzian signature $(+,-,-,-)~$ and on the boundary $\partial \mathcal{T}$, $L$ can be continuously extended as $L|_{\partial \mathcal{T}} = 0$. \footnote{It is possible to equivalently formulate this property with opposite sign of $L$ and metric $g^L$ of signature $(-,+,+,+)$. We fixed the signature and sign of $L$ here to simplify the discussion.}
\end{itemize}

This is a refined version of the definition of Finsler spacetimes in Reference~\cite{Hohmann:2018rpp} and basically covers, if~one chooses $\mathcal{A} = \mathcal{T}$, the~improper Finsler spacetimes defined in Reference~\cite{Bernal:2020bul}.

The $1$-homogeneous function $F$, which defines the point particle action for curves $\gamma$ on $M$\,,
\begin{align}
S[\gamma] = \int d\tau\ F(\gamma,\tfrac{d\gamma}{d\tau})\,,
\end{align}
is obtained from the Finsler-Lagrange function $L$ as $F=\sqrt{|L|}$ and interpreted as proper time integral of observers. For~clarity, we list the different sets which appear in the above definition and comment on their~meaning:
\begin{itemize}[leftmargin=2.3em,labelsep=5.8mm]
	\item $\mathcal{A}$: the subbundle where $L$ is defined, smooth and $g^L$ is nondegenerate, with~fiber $\mathcal{A}_{x} = \mathcal{A} \cap T_xM$, called the set of \emph{admissible vectors};
	\item $\mathcal{T}$: the set of future pointing timelike directions, a~maximally connected conic subbundle where $L > 0$, the~$L$-metric exists and has Lorentzian signature $(+,-,-,-)$, with~fiber $\mathcal{T}_x = \mathcal{T} \cap T_xM$;
	\item $\mathcal{N} = \{(x,\dot x)| L(x,\dot x)=0\}$: the subbundle where $L=0$, with~fiber $\mathcal{N}_x = \mathcal{N} \cap T_xM$.
\end{itemize}
We like to point again the relation $\partial \mathcal{T}\subset \mathcal{N}$ is demanded in our~definition. 

An important building block of the geometry of Finsler spacetimes is the geodesic spray, locally~given by the coefficients
\begin{align}
G^a = \frac{1}{4}g^{Laq}(\dot x^m \partial_m \dot \partial_q L - \partial_q L)\,.
\end{align}

It defines the Finsler geodesic equation in arclength parametrization $\ddot x^a + 2 G^a(x,\dot x)=0$, the~canonical (Cartan) nonlinear connection coefficients $G^a{}_b = \dot \partial_b G^a$ and the Berwald linear connection coefficients $G^a{}_{bc} = \dot \partial_c\dot \partial_b G^a$.

A Finsler spacetime is called of Berwald type~\cite{Berwald1926,Bao}, or~simply Berwald spacetime, if~and only if, in~any local chart, the~geodesic spray is quadratic in its dependence on the tangent space coordinates $\dot x$:
\begin{align}\label{eq:BerwG}
G^a(x,\dot x) = \frac{1}{2}G^a{}_{bc}(x)\dot x^b \dot x^c\,.
\end{align}

This is equivalent to demanding that the canonical nonlinear connection coefficients are actually linear in their $\dot x$ dependence, or~that the Berwald linear connection coefficients are independent of $\dot x$. The~latter means that the $G^a{}_{bc}(x)$ define an affine connection on $M$.

Next, we will determine the most general cosmologically symmetric Finsler spacetimes from the CP, before~we derive the most general homogeneous and isotropic Berwald spacetime in Section~\ref{sec:BerwCond}.

%%%%%%%%%%%%%%%%%%%%%%%%%%%%%%%%%%%%%%%%%%%%%%%%%%%%%%%%%%%%%%%%%%%%%%%%%%%%%%%%%%%%%%%
\section{The Cosmological Principle on Finsler~Spacetimes}\label{ssec:cosmo}

We will prove in the following that, applying the CP to Finsler spacetime geometry singles out 6~symmetry generating vector fields, which characterize spatially homogeneous and isotropic Finsler spacetimes. More precisely, these symmetry generating vector fields are the same as the ones defining cosmological symmetry in general relativity, that is, the~same as in the case of pseudo-Riemannian geometry. The~key argument is that spatial homogeneity and isotropy ensure the existence of a maximally symmetric Riemannian metric on the $3$-dimensional time slice hypersurfaces of a Finsler spacetime. The~Finsler Lagrangians we will determine will involve this maximally symmetric Riemannian metric, albeit, in~a non-trivial~way.

We begin by assuming that the spacetime in consideration possesses a smooth global time function $t:M\rightarrow \mathbb{R}$, which assigns to each $p \in M$ a time stamp and whose differential $dt$ satisfies $dt(X)>0,\ \forall X\in \mathcal{T} \cup \partial \mathcal{T}$. The~level sets of the time function $\Sigma_T=\{p\in M| t(p)= T = constant\}$ are interpreted as equal-time spatial hypersurfaces. We assume that these level sets are~connected.

The demand of the existence of a symmetry group acting on the hypersurfaces $\Sigma_T$ is the core of the cosmological~principle.

\subsection{The Symmetry and the Isotropy~Group}
A Finsler spacetime geometry satisfies the CP if it is (see for example Wald~\cite{Wald} or Weinberg~\cite{Weinberg} for pseudo-Riemannian spacetimes):
\begin{itemize}[leftmargin=2.3em,labelsep=5.8mm]
	\item \emph{spatially homogeneous}, that is,\ for each fixed value $T$ of the time function $t$ and for any two points $q_1$ and $q_2$ in $\Sigma_T$, there exists a diffeomorphism of $M$ which maps $q_1$ to $q_2$ and preserves the Finsler Lagrangian $L$.  In~other words, there exists a Lie Group G of isometries of $(M,L)$ acting transitively on each slice \(\Sigma_T\);   % Is the italics necessary?
	\item \emph{spatially isotropic}, that is,\ around each $p\in M$ there exists a congruence of observer curves $\gamma(t,s)$ with tangent vector field $\frac{\partial \gamma}{\partial t} \in \mathcal{T}$, such that for each two lines $[Z_1]$ and $[Z_2]$ in $T_p\Sigma_T$ there exists a diffeomorphism of $M$ preserving $p$, $\frac{\partial \gamma}{\partial t}$ and $L$ while mapping $[Z_1]$ to $[Z_2]$ (the equivalence relation $[]$ is defined as $Z \sim \bar Z$ if $Z = \lambda Z, \lambda \in R^*$). In~other words, the~stabilizer $G_p = \{\psi \in G| \psi(p) = p\}$ at $p\in M$ acts transitively on the projective tangent spaces $PT_p\Sigma_T$ of $\Sigma_T$.
\end{itemize}

Fix $T \in \mathbb{R}$. We start by a first remark: if a Lie group acts transitively on a connected manifold, then~the connected component of the identity element is a Lie subgroup that still acts transitively on the respective manifold~\cite{Onishchik} (p. 398).

Since the slice $\Sigma_T$ is connected, there is no loss of generality if we assume that $G$ is connected (in the contrary case, we can restrict our attention to the connected component $G_0$ of the identity in $G$, which still acts transitively on $\Sigma_T$. The~corresponding isotropy group will then be $G_p \cap G_0$). The~connected component of the identity in $G$ only contains orientation-preserving diffeomorphisms $\psi: M \rightarrow M$.

We note that the homogeneity demand makes $(\Sigma_T, G)$ a homogeneous space; thus, $\Sigma_T$ is diffeomorphic to the quotient $G/G_p$ and the following relation between the dimensions of the involved sets holds
\begin{align}
	\dim G = \dim G_p + \dim \Sigma_T = \dim G_p + 3.
\end{align}

We will now prove two lemmas to identify the dimension of the groups $G$ and $G_p$.
\begin{Lemma}\label{lem1}
	On a Finsler spacetime $(M,L)$ satisfying the CP, the~dimension of the space of Killing vectors at each point $p\in M$ is at most 6 and the dimension of the isotropy group $G_p$ is at most 3.
\end{Lemma}
\begin{proof}
	Locally, in~a coordinate chart around each $p\in M$, the~action of $G$ on $M$ is determined by the generating vector fields $\xi=\xi^a\partial_a$; the 1-parameter subgroup of $G$ generated by such a vector field $\xi$ acts on $M$ by some diffeomorphisms $\psi$ locally expressed as: $\psi^a(x) = x^a + \epsilon \xi^a(x) +\mathcal{O}(\epsilon^2)$. On~a Finsler spacetime $(M,L)$, these diffeomorphsims are isometries if and only if $\xi^C(L)=0$, where~\mbox{$\xi^C = \xi^a \partial_a + \dot x^b \partial_b \xi^a \dot \partial_a$} is the complete lift of $\xi$ to the tangent~bundle.

	Using the Killing equation $\xi^C(L)=0$ on a Finsler spacetime $(M,L)$, it was proven in Reference~\cite{Pfeifer:2013gha} (Section 5.4.4) that any Killing vector field $\xi=\xi^a \partial_a$ in local coordinates around any point $p\in M$, is~locally uniquely determined by the values $\xi^a(p)$ and by the derivatives $\partial_b \xi^a(p)$, with~$a<b$. For~an $n$-dimensional Finsler spacetime this means, we can freely choose at most $n$ values  $\xi^a(p)$ and at most $\frac{n(n-1)}{2}$ values $\partial_b\xi^a(p)$; all in all, this gives at most $n+ \frac{n(n-1)}{2}= \frac{n(n+1)}{2}$ independent Killing vectors. Thus, for~$n=4$ the dimension of the isometry group can be at most~10.

	Choose an arbitrary local chart such that the first coordinate is the time function $x^0=t$. Since~the diffeomorphisms $\psi\in G$ preserve the time slices $\Sigma_T$, we find that $\xi(T) = 0$, that is,\ $\xi^0$ and its derivatives identically vanish and that the remaining components $\xi^i, i=1,~2,~3$ can only depend on $x^i, i=1,2,3$. That is, we can only pick freely, at~a given point, 3 values $\xi^a(p)$ and 3 derivatives $\partial_b\xi^a(p)$, with~$a<b$\change{,} which gives us at most 6 degrees of freedom for the space of Killing vectors at each point. In~particular, since the isotropy subgroup $G_p$ fixes the point $p$, we must have $\xi^a(p)=0$ for all $a=0,\ldots,3$, hence we can only freely choose the 3 derivatives $\partial_b\xi^a(p)$ with $a<b$. Consequently, the~dimension of $G_p$ cannot exceed 3, which proves our first remark.
\end{proof}

\begin{Lemma}\label{lem2}
	The dimension of the isotropy group $G_p$ is at least 3.
\end{Lemma}
\begin{proof}
	Let us properly understand the statement ``the isotropy group $G_p = \{\psi \in G| \psi(p) = p\}$ at $p\in M$ acts transitively on the projective tangent spaces $PT_p\Sigma_T$ of $\Sigma_T$''. We will identify a transitive and \emph{effective} action of a quotient group of $G_p$ on the projective tangent space $PT_p\Sigma_T$. This group is obtained by identifying elements of $G_p$ that have the same linear tangent map at $p$.

	Any Lie group action on a manifold gives rise to an effective Lie group action on the respective manifold, by~factorizing away the elements that provide trivial actions. In~our case, assume the group $G_p$ does not act effectively on the (for the moment, non-projectivized) tangent space $T_p \Sigma_T$ and denote by $Id_p$ the subset of $G_p$ whose elements provide trivial actions. Then, $Id_p$ is a normal subgroup of $G_p$ and the factor group
\begin{align}
		G'_p = G_p/Id_p
	\end{align}
	acts effectively on $T_p \Sigma_T$, by~the rule $(d\psi \circ Id_p)v := d\psi\ (v) $.

	The group $G'_p$ can be identified with a subgroup of the general linear group $GL_3(\mathbb{R})$ (more precisely, of~the connected component of $GL_3(\mathbb{R})$ consisting with matrices with positive determinants). This is justified as follows. Fix an arbitrary coordinate chart on $TM$. The~action the subgroup $G_p$ fixing $p$ on the tangent space $T_p \Sigma_T$ is then expressed as a matrix multiplication\change{,} namely, by~the Jacobian matrices of the diffeomorphisms $\psi \in G_p$. Factorizing $G_p$ by the subgroup $Id_p$ actually means identifying as a single element those diffeomorphisms $\psi, \psi'$ of $M$ which have the same values at $p$ and the same Jacobian matrices (but whose higher order derivatives at $p$ might differ). This way, the~mapping from $G'_p$ to $GL_3(\mathbb{R})$ associating to a class $[\psi] \in G'_p$ the Jacobian matrix at $p$ of $\psi$, is an injective homomorphism, which gives us the right of identifying $G'_p$ as a subgroup of $GL_3(\mathbb{R})$.

	Further, passing to the projectivised tangent spaces (that can be identified with the projective plane $P \mathbb{R}^3$ once a choice of the basis of $T_p \Sigma_T$ is made), the~group that naturally acts is then the subgroup $PG'_p$ of the projective group  $PGL_3(\mathbb{R})$  obtained by factorizing $GL_3(\mathbb{R})$ by the group of rescalings $A \mapsto \lambda A$, with~$\lambda \in \mathbb{R}$.
	Hence, on~one hand, the~isotropy request  that $G_p$ acts transitively on the projective tangent space $PT_p \Sigma_T$ implies that the group $PG'_p$ (which is isomorphic to a subgroup of the projective linear group $PGL_3(\mathbb{R})$), acts transitively and effectively on $PT_p \Sigma_T$. The~latter is true since $PGL_3(\mathbb{R})$ acts effectively on $PT_p \Sigma_T$ and so do all of its subgroups, hence in particular $PG'_p$. On~the other hand, we~notice that the only rescalings that can belong to $G'_p$ are those with positive factors~$\lambda$. But~these cannot be $L$-isometries unless $\lambda = 1$, since $L(\lambda V) = \lambda^2 v$ for all $v \in T_p \Sigma_T$. That is, the~subgroup of $G'_p$ we have used for factorization is actually trivial, that is,
\begin{align}
	G'_p \simeq PG'_p.
	\end{align}
	
There is actually a smaller subgroup of $PGL_3(\mathbb{R})$ which still acts transitively on  $PT_p \Sigma_T$. To~identify this subgroup, we mention the following result (\cite{Montgomery}, \cite{Onishchik}, p. 398):
 	If a connected Lie group acts on a compact manifold with finite fundamental group, then any maximal compact subgroup also acts transitively on the respective~manifold.

	The projectivized space $PT_p \Sigma_T$ is connected, compact and its fundamental group is $\mathbb{Z}_2$. Since~$PT_p \Sigma_T$ is connected, the~connected component of the identity in $G'_p$ still acts transitively and effectively on it. Denoting by $H_p$ a maximal compact subgroup of the connected component of the identity in $PG'_p$, we find that $H_p$ also acts transitively and effectively on~$PT_p \Sigma_T$.

	Finally, we use a result in Reference~\cite{Onishchik} (pp. 398--401), stating that any connected, compact Lie group acting transitively and effectively on the projective plane $P\mathbb{R}^3$ is isomorphic to $SO(3)$; that is, our subgroup $H_p$ of the isotropy group $PG_p'$ must be isomorphic to $SO(3)$ and thus:
\begin{align}
		\dim G_p \geq \dim G'_p = \dim PG'_p \geq \dim H_p = 3.
	\end{align}
\end{proof}

From Lemma \ref{lem1} and Lemma \ref{lem2} we find
\begin{Theorem}
On a Finsler spacetime satisfying the cosmological principle, the~dimension $\dim G_p$ of the isotropy group $G_p$ is $3$ and the dimension $\dim G$ of the full symmetry group $G$ is $6$.
\end{Theorem}

\begin{Remark}
The above result intuitively tells us that, on~one hand, by~taking into account only the linear approximation at $p$ of the diffeomorphisms $\psi$ and, on~the other hand, by~taking into account just a bounded region around the identity of the symmetry group, we do not lose any generators of the Lie algebra.
\end{Remark}

We can actually state a much stronger~result.

\begin{Remark}
	On Finsler spacetimes with cosmological symmetry, the~isotropy group $G_p$ is compact. 
\end{Remark}
	
\begin{proof}
	First we show that $G_p$ is connected, and~then we apply Cartan's Theorem on connected Lie~groups.
	\begin{itemize}[leftmargin=2.3em,labelsep=5.8mm]
	\item Under the above assumption that $G$ is connected, the~isotropy group $G_p$ is also connected. To~see this, let us assume that $G_p$ is the disjoint union of at least two connected components, say,~$G_p = A \sqcup B$, where $A$ is the component of the identity. Since $PT_p \Sigma_T$ is connected and $G_p$ acts transitively on it, it follows that $A$ also acts transitively on $PT_p \Sigma_T$ and see Reference~\cite{Onishchik} (p. 395),
\begin{align}
		G_p = AH_v,
	\end{align}
	where $H_v$ is the stabilizer of a point $[v]$ in 	$PT_p \Sigma_T$. But, $H_v = \{\psi \in G_p | [d\psi\ v]=[v], \forall v \in T_p \Sigma_T \}$ consists of rescalings with positive factors; recalling that rescalings with positive factors cannot be $L$-isometries, it follows that $H_v$ is trivial, hence, $G_p$ coincides with its identity connected component $A$.
	\item Further, we apply Cartan's classification theorem (\cite{Onishchik}, p. 389 ), to~$G_p$. The~Theorem states that any connected Lie group is the direct product between one of its maximal compact subgroups, say $\mathcal{H}$, and~a Euclidean space. Since $G_p$ acts transitively on the compact manifold $PT_p \Sigma_T$ (which, as~we have seen above, has finite fundamental group), its maximally compact subgroup $\mathcal{H}$ also acts transitively on $PT_p \Sigma_T$. But, the~smallest compact group that can act transitively on $PT_p \Sigma_T$ is 3-dimensional (more precisely, $SO(3)$), that is, $\dim \mathcal{H} \geq 3$. Taking into account that the dimension of $G_p$ itself is 3, it means that $G_p = \mathcal{H}$, that is $G_p$ itself is compact.
\end{itemize}
\end{proof}	

\subsection{The Symmetry~Generators}
To explicitly determine the generators of the groups $G$ and $G_p$, we use the above remark, which~states that $G_p$ is compact. Then, we take into account that $\Sigma_T$ is a homogeneous manifold having a compact isotropy group and thus it must admit a $G$-invariant Riemannian metric $h$ (\cite{KobayashiNomizu1}, Example 1.3, p. 154). In~other words, the~generators of our group $G$ are also Killing vector fields of $h$. But, for~a $3$ dimensional Riemannian manifold there can exist at most $6$ Killing vector fields, hence~$\Sigma_T$ is maximally symmetric and $h$ is the corresponding maximally symmetric metric. In~particular, $h$~has constant scalar curvature $k$. The~Riemannian metric $h$ also gives us the possibility of having local spherical coordinates $(t, r,\theta,\phi)$ around $p$. In~these coordinates, see Reference~\cite{Weinberg,Pfeifer:2011xi,Garecki}, the~isotropy group generators (which are the elements of the Lie algebra $so(3)$) are written as:
\begin{align}\label{eq:iso}
	X_1 &= \sin\phi\partial_{\theta} + \cot\theta \cos\phi \partial_{\phi},\quad X_2 = -\cos\phi\partial_{\theta} + \cot\theta \sin\phi\partial_{\phi},\quad X_3 = \partial_\phi \,,
\end{align}
and the generators of quasi translations, as:
\begin{equation}\label{eq:trans}
\begin{aligned}
	X_4 &= \sqrt{1-k r^2} \sin\theta \cos\phi \partial_r + \frac{\sqrt{1-k r^2}}{r}\cos\theta\cos\phi \partial_{\theta} - \frac{\sqrt{1-k r^2}}{r} \frac{\sin\phi}{\sin\theta}\partial_{\phi}\\
	X_5 &= \sqrt{1-k r^2} \sin\theta \sin\phi \partial_r + \frac{\sqrt{1-k r^2}}{r}\cos\theta\sin\phi \partial_{\theta} + \frac{\sqrt{1-k r^2}}{r} \frac{\cos\phi}{\sin\theta}\partial_{\phi}\\
	X_6 &= \sqrt{1-k r^2} \cos\theta \partial_r - \frac{\sqrt{1-k r^2}}{r}\sin\theta \partial_{\theta}\,.
\end{aligned}
\end{equation}

Finally, solving the Finsler Killing equation $X^C_I(L)=, I=1,...,6$, where $X^C= X^a \partial_a + \dot x^b \partial_b X^a \dot \partial_a$ yields that
\begin{align}
	L(t,r,\theta,\phi,\dot t,\dot r,\dot \theta,\dot \phi) = L(t,\dot t,w), \quad w^2 = \frac{\dot r^2}{1- kr^2} + r^2 (\dot \theta^2 + \sin^2\theta \dot\phi^2)\,,
\end{align}
is the most general spatially homogeneous and isotropic Finsler Lagrangian. The~explicit calculation, as~well as the expressions for the complete lifts $X^C_I$ can be found for example in Reference~\cite{Pfeifer:2011xi}.

%%%%%%%%%%%%%%%%%%%%%%%%%%%%%%%%%%%%%%%%%%%%%%%%%%%%%%%%%%%%%%%%%%%%%%%%%%%%%%%%%%%%%%%
\section{Homogeneous and Isotropic Berwald~Spacetimes}\label{sec:BerwCond}
In order to find the desired Berwald Finsler spacetimes, we rewrite a generic Finsler Lagrangian in a specific way, which allows us to reduce the condition that a Finsler spacetime shall be Berwald, to~a first order partial differential~equation.

\subsection{The Berwald~Condition}
Every Finsler spacetime Lagragian $L$ can be written as $L(x,\dot x) = g(\dot x, \dot x)\Omega(x,\dot x)$, where $g$ is an arbitrary pseudo-Riemannian metric, $g(\dot x, \dot x) = g_{ab}(x)\dot x^a \dot x^b$ and $\Omega=\Omega(x,\dot x)$ is a $0$-homogeneous function in $\dot x$. In~Reference~\cite{Pfeifer:2019tyy}, it was proven that $L$ defines a Berwald Finsler geometry if and only if there exist a  $g$ and a $(1,2)$-tensor field $D$ on $M$, symmetric in its vector arguments, such that $\Omega$ satisfies the equation
\begin{align}\label{eq:berwcond}
\partial_a \Omega(x,\dot x) - \Gamma^b{}_{ac}(x)\dot x^c \dot \partial_b \Omega(x,\dot x) = D^b{}_{ac}(x)\dot x^c \left(\dot \partial_b\Omega + \frac{2\dot x_b \Omega(x,\dot x)}{g(\dot x, \dot x)}\right)\,,
\end{align}
which we call \emph{the Berwald condition}. %Is the italics necessary
Here, the~indices were raised and lowered with the pseudo-Riemannian metric, that is, $\dot x_b = \dot x^a g_{ab}(x)$. The~connection coefficients $G^a{}_{bc}$ in \eqref{eq:BerwG} are then determined by the Christoffel symbols $\Gamma$ of $g$ and the tensor $D$ as
\begin{align}\label{eq:affcon}
G^a{}_{bc} = \Gamma^a{}_{bc} + D^a{}_{bc}\,.
\end{align}

In case the expansion of $L = g(\dot x, \dot x)\Omega(x,\dot x)$ is done with a proper pseudo-Riemannian metric, where~$g(\dot x,\dot x)$ has a non trivial null structure, one might worry what happens on this null~structure.

To avoid this problem observe that, if~one has found one pseudo-Riemannian metric $g$, a~tensor $D$ and a factor $\Omega$ which solve \eqref{eq:berwcond}, the~Berwald condition is satisfied for any alternative expansion of $L$ of this type, that is, for~any other pseudo-Riemannian metric $\tilde g$, we can find a corresponding (1,2)-tensor field $\tilde D$ and a factor $\tilde \Omega$.

To see this, let us expand $L=g(\dot x,\dot x)\Omega(x,\dot x) = \tilde g(\dot x,\dot x) \tilde \Omega(x,\dot x)$, where $g$ and $\Omega$ satisfy \eqref{eq:berwcond} for a tensor field $D$. Then $\delta_a L = \partial_a - G^b{}_a\dot{\partial}_bL=0$ implies \vspace{6pt}
\begin{align}
	0 = (\partial_a \tilde \Omega - G^b{}_a \dot{\partial}_b\tilde \Omega)\tilde g(\dot x,\dot x) + 2 \tilde x_c (\tilde \Gamma^c{}_{ab}\dot x^b - G^c{}_a)\tilde \Omega\,.
\end{align}

Using that $G^a{}_b = (\Gamma^a{}_{bc} + D^a{}_{bc})\dot x^c$ and introducing the tensor field components $\tilde D^a{}_{bc} = D^a{}_{bc} + \Gamma^a{}_{bc} - \tilde \Gamma^a{}_{bc}$ we find
\begin{align}
	\partial_a \tilde \Omega(x,\dot x) - \tilde \Gamma^b{}_{ac}(x)\dot x^c \dot \partial_b \tilde\Omega(x,\dot x) =\tilde D^b{}_{ac}(x)\dot x^c \left(\dot \partial_b\tilde \Omega + \frac{2\dot x_b \tilde \Omega(x,\dot x)}{\tilde g(\dot x, \dot x)}\right)\,.
\end{align}

Hence, also $\tilde g$, $\tilde \Omega$ and $\tilde D$ satisfy the Berwald condition. Thus the metric factor in the expansion of the Finsler Lagrangian can be chosen arbitrarily; in particular, we can choose a positive definite metric $g$ in order to avoid complications on the null structure of the~metric.

In the following, we will insert the most general $g$ and $D$ which are compatible with a specific spacetime symmetry and solve the Berwald condition for $\Omega$. This determines the most general Berwald Finsler spacetime for the desired spacetime~symmetry.

\subsubsection{The Cosmological Berwald~Condition}
To evaluate the Berwald condition, we consider Finsler Lagrangians $L$ which are the form $L(t,\dot t, w) = (\dot t^2 +\sigma a(t)^2 w^2)\Omega(t,\dot t, w)$, where the metric factor $g(\dot x,\dot x)=(\dot t^2 + \sigma a(t)^2 w^2)$ is the most general homogeneous and isotropic (pseudo)-Riemannian metric with positive definite ($\sigma=1$) resp. Lorentzian ($\sigma=-1$) signature. As~we discussed, the~Berwald condition \eqref{eq:berwcond} does not depend on this choice, which will also explicitly follow from the upcoming calculation. The~possibility of choosing $\sigma=1$ means that the set of null vectors of the chosen metric does not interfere with our result. On~the conic bundle $\mathcal{T}$, where $\dot t\neq 0$, this expression can be nicely rewritten in terms of the $0$-homogeneous variable $s=w/\dot t$
\begin{align}\label{eq:Lcosmo}
L(t,\dot t, w) = \dot t^2 (1 + \sigma a(t)^2 s^2)\Omega(t,1,s)\,,
\end{align}
which will be very convenient to evaluate the Berwald~condition.

The second ingredient in this condition is the $(1,2)$-tensor field $D$. The~most general spatially homogeneous and isotropic such tensor field that is symmetric in its vector arguments, has the following nonzero components, see for example Reference~\cite{Hohmann:2019fvf},
\begin{align}\label{eq:Tcosmo}
D^t{}_{tt} &= b(t),\ D^t{}_{rr} = \frac{c(t)}{1-kr^2},\ D^t{}_{\theta\theta} = r^2 c(t),\ D^t{}_{\phi\phi} = r^2 \sin^2\theta c(t)\,,\\
D^r{}_{rt} &= D^r{}_{tr} = D^\theta{}_{\theta t} = D^\theta{}_{t\theta} = D^\phi{}_{\phi t} = D^\phi{}_{t\phi} = d(t)\,,
\end{align}
where $b(t), c(t), d(t)$ are arbitrary functions of $t$.

Using the above expressions in the Berwald condition \eqref{eq:berwcond} yields two independent equations, which need to be solved to determine $\Omega(t,s) \equiv \Omega(t,1,s)$. Let $'$ denote derivatives with respect to the single argument of the functions $a,b,c$ and $d$, then the spatial equations, if~the index $a$ assumes the values $r,\theta$ or $\phi$, read
\begin{align}\label{eq:BCspat}
2\frac{c(t) + \sigma a(t)^2 d(t)}{1 + \sigma s^2 a(t)^2}\Omega(t,s) + \frac{a'(t) - a(t)\big[s^2(c(t) - \sigma a(t)a'(t))-d(t)\big]}{s a(t)}\frac{\partial}{\partial s}\Omega(t,s) = 0
\end{align}
and the temporal equation, for~which the index $a$ assumes the value $t$ is
\begin{align}\label{eq:BCtemp}
\frac{\partial}{\partial t}\Omega(t,s) - 2\frac{b(t) + \sigma s^2 a(t)^2 d(t)}{1 + \sigma s^2 a(t)^2}\Omega(t,s) + \frac{a(t)\big[b(t)-d(t)\big]-a'(t)}{a(t)}s\frac{\partial}{\partial s}\Omega(t,s)= 0.
\end{align}

%%%%%%%%%%%%%%%%%%%%%%%%%%%%%%%%%%%%%%%%%%%%%%%%%%%%%%%%%%%%%%%%%%%%%%%%%%%%%%%%%%%%%%%
\subsection{Solving the Cosmological Berwald~Condition}\label{ssec:cosmosol}
Solving the Berwald condition for the most general combination of $\Omega(t,s)$, $b(t),c(t)$ and $d(t)$ is not trivial. We will now classify all the solutions of the system \eqref{eq:BCspat} and \eqref{eq:BCtemp}. 

Introducing the functions
\begin{align}
M(t,s) &= 2\frac{c(t) + \sigma a(t)^2 d(t)}{1 + \sigma s^2 a(t)^2}, \quad N(t,s) = \frac{a'(t) - a(t)\big[s^2(c(t) - \sigma a(t)a'(t))-d(t)\big]}{s a(t)}\\
P(t,s) &= - 2\frac{b(t) + \sigma s^2 a(t)^2 d(t)}{1 + \sigma s^2 a(t)^2} \quad Q(t,s) = \frac{a(t)\big[b(t)-d(t)\big]-a'(t)}{a(t)}s\,,
\end{align}
the Berwald condition becomes
\begin{align}\label{eq:BerwC2}
M(t,s) \Omega(t,s) + N(t, s)\frac{\partial}{\partial s}\Omega(t,s) = 0,\quad  \frac{\partial}{\partial t}\Omega(t,s) +P(t,s) \Omega(t,s) + Q(t,s)\frac{\partial}{\partial s}\Omega(t,s)= 0\,.
\end{align}

We now analyze the first equation and find several cases in which we obtain trivial solutions, in~the sense that the Finsler Lagrangian is  pseudo-Riemannian or zero. The~only case that provides proper Finslerian solutions is $M=N=0$, as~we will see~below.

%%%%%%%%%%%%%%%%%%%%%%%%%%%%%%%%%%%%%%%%%%%%%%%%%%%%%%%%%%%%%%%%%%%%%%%%%%%%%%%%%%%%%%%
\subsubsection{Trivial~Solutions}\label{ssec:MNnonzero}
Trivial solutions arise in the following~situations:
\begin{itemize}[leftmargin=2.3em,labelsep=5.8mm]
\item
If $N$ is different from zero, then we can divide the first equation in \eqref{eq:BerwC2} by \(N\),
\begin{align}\label{eq:MNtrivial}
\frac{M(t,s)}{N(t,s)}\Omega(t,s) + \frac{\partial}{\partial s}\Omega(t,s) = 0\,.
\end{align}
Now it is helpful to introduce the function
\begin{align}
A(t, s) = \frac{1 + \sigma s^2 a(t)^2}{a'(t) - a(t)\big[s^2(c(t) - \sigma a(t)a'(t))-d(t)\big]}\,
\end{align}
and to realize that it satisfies
\begin{align}
\frac{1}{A(t, s)}\frac{\partial}{\partial s}A(t, s) = \frac{M(t,s)}{N(t,s)}\,.
\end{align}
%\begin{align}
%\frac{1}{A(t, s)}\frac{\partial}{\partial s}A(t, s) = \frac{a(t)^2d(t) - c(t)}{a(t)\big[d(t) - s^2 (c(t) + a(t)a'(t))\big] + a'(t)} \times \frac{2sa(t)}{s^2a(t)^2 - 1} = \frac{M(t,s)}{N(t,s)}\,.
%\end{align}
We can thus rewrite Equation~\eqref{eq:MNtrivial} as
\begin{align}
\frac{\Omega(t, s)}{A(t, s)}\frac{\partial}{\partial s}A(t, s) + \frac{\partial}{\partial s}\Omega(t,s) = 0\,.
\end{align}
After multiplication with \(A(t, s)\) we find
\begin{align}
\frac{\partial}{\partial s}\left(\Omega(t, s)A(t, s)\right) = 0 \quad \Rightarrow \quad \Omega(t, s)A(t, s) = f(t)\,,
\end{align}
where $f(t)$ is an arbitrary function of $t$. Using the explicit form of the function $A$, we thus obtain the solution
\begin{align}\label{eq:omegaspat}
\Omega(t,s) = \frac{f(t)}{A(t, s)} = f(t)\frac{a'(t) - a(t)\big[s^2(c(t) - \sigma a(t)a'(t))-d(t)\big]}{1 + \sigma s^2 a(t)^2} \,.
\end{align}
Constructing the Finsler Lagrangian $L = \dot t^2(1-a(t)^2 s^2)\Omega(t,s)$ from this solution immediately~yields \vspace{6pt}
\begin{align}
L = I(t) \dot t^2 + J(t) w^2\,,
\end{align}
with $I(t) = f(t) ( a'(t) + a(t) d(t))$ and $J(t) = a(t) f(t) ( \sigma a(t) a'(t) - c(t) )$. The~just constructed Finsler Lagrangian is again quadratic in its dependence on the velocities and hence defines a pseudo-Riemannian spacetime geometry. In~the particular case when $M(t,s)$ is zero, we see from~\eqref{eq:MNtrivial} that $\Omega$ is just a function of $t$.

\item
If $N = 0$ and $M\neq 0$, then the first equation in \eqref{eq:BerwC2} implies immediately that $\Omega(t,s) = 0$ and thus the Finsler Lagrangian is $L = 0$.
\end{itemize}

From this analysis, we find that nontrivial cosmologically symmetric Berwald Finsler Lagrangians can only be obtained if $M=N=0$.

%%%%%%%%%%%%%%%%%%%%%%%%%%%%%%%%%%%%%%%%%%%%%%%%%%%%%%%%%%%%%%%%%%%%%%%%%%%%%%%%%%%%%%%
\subsubsection{Finslerian~Solutions}\label{sssec:cosmosolfins}
Demanding that $M=N=0$ leads to the equations
\begin{align}
c(t) + \sigma a(t)^2 d(t) = 0, \quad a'(t) - a(t)\big[s^2(c(t) - \sigma a(t)a'(t))-d(t)\big]=0
\end{align}
for $c(t)$ and $d(t)$. Since $s$ and $t$ are independent variables, the~$s$ dependence in the latter must vanish, which immediately implies $c(t) = \sigma a(t)a'(t)$. Plugging this into the remaining equations yields $d(t) = - \frac{a'(t)}{a(t)}$.

Having solved \eqref{eq:BCspat} for general $\Omega(t,s)$, the~remaining Equation  \eqref{eq:BCtemp} becomes
\begin{align}\label{eq:BCnontriv}
\frac{\partial}{\partial t}\Omega(t,s) + s b(t) \frac{\partial}{\partial s}\Omega(t,s) - \frac{2( b(t) - \sigma s^2 a(t) a'(t) ) }{1 + \sigma s^2 a(t)^2} \Omega(t,s) = 0\,.
\end{align}
%\begin{align}\label{eq:BCnontriv}
%\frac{\partial}{\partial t}\Omega(t,s) + s b(t) \frac{\partial}{\partial s}\Omega(t,s) + \frac{2( b(t) + s^2 a(t) a'(t) ) }{s^2 a(t)^2 - 1} \Omega(t,s) = 0\,.
%\end{align}

This equation intertwines the $t$ and $s$ dependence of $\Omega$ and can be solved by a change of variables. For~this purpose we substitute \(s\) by a new variable \(u\), which is defined such that
\begin{align}
s = uB(t)\,, \quad B'(t) = B(t)b(t)\,.
\end{align}

In other words, the~function \(B(t)\) is given explicitly as the integral
\begin{align}\label{eq:Bint}
B(t) = \exp\left(\int_{t_0}^t b(\tau)d\tau\right)\,,
\end{align}
up to an undetermined constant of integration related to the choice of the lower bound \(t_0\). Replacing~\(\Omega(t, s)\)~by
\begin{align}
\tilde{\Omega}(t, u) = \Omega(t, s) = \Omega(t, uB(t))\,,
\end{align}
we find that
\begin{align}
\frac{\partial}{\partial u}\tilde{\Omega}(t, u) = B(t)\frac{\partial}{\partial s}\Omega(t, s)\,, \quad
\frac{\partial}{\partial t}\tilde{\Omega}(t, u) = \frac{\partial}{\partial t}\Omega(t, s) + uB'(t)\frac{\partial}{\partial s}\Omega(t, s)\,.
\end{align}

Using
\begin{align}
uB'(u) = uB(t)b(t) = sb(t)\,,
\end{align}
our original Equation~\eqref{eq:BCnontriv} becomes
\begin{align}\label{eq:BCnontriv2}
\frac{\partial}{\partial t}\tilde{\Omega}(t, u) - 2\frac{B'(t)/B(t) - \sigma u^2 B(t)^2 a(t)a'(t) }{1 + \sigma u^2B(t)^2a(t)^2 }\tilde{\Omega}(t, u) = 0\,,
\end{align}
and so it contains derivatives with respect to \(t\) only. We can explicitly integrate this equation by introducing another function
\begin{align}
C(t, u) = u^2a(t)^2 + \frac{\sigma}{B(t)^2} \quad \Rightarrow \quad \frac{\partial}{\partial t}C(t, u) = 2u^2a(t)a'(t) + 2\frac{B'(t)}{B(t)^3}\,,
\end{align}
hence simplifying Equation~\eqref{eq:BCnontriv2} to
\begin{align}
\frac{\partial}{\partial t}\tilde{\Omega}(t, u) + \frac{\tilde{\Omega}(t, u)}{C(t, u)}\frac{\partial}{\partial t}C(t, u) = 0\,.
\end{align}

After multiplication by \(C(t, u)\) we thus conclude
\begin{align}
\frac{\partial}{\partial t}\left(\tilde{\Omega}(t, u)C(t, u)\right) = 0 \quad \Rightarrow \quad \tilde{\Omega}(t, u)C(t, u) = f(u)\,,
\end{align}
with an arbitrary free function \(f(u)\) which depends only on \(u\). Substituting back we obtain the general Finslerian solution of the Berwald condition
\begin{align}
\Omega(t,s) = \frac{B(t)^2}{s^2a(t)^2 + \sigma}f(sB(t)^{-1})\,.
\end{align}

Recall from the definition~\eqref{eq:Bint} that \(B(t)\) is determined only up to a multiplicative constant; however, this constant can simply be absorbed into the function \(f\). With~this result, we found the most general nontrivial Berwald Finsler spacetimes with cosmological symmetry:
\begin{align}
L = \dot{t}^2 (1 + \sigma a(t)^2 s^2) \Omega(t,s) = \sigma \dot{t}^2B(t)^2f(sB(t)^{-1})\,.
\end{align}

The only remnant of the two different possible choices for the metric in the decomposition of the Finsler Lagrangian appears in the overall factor $\sigma$, which can, without~loss of generality, be absorbed into the free function $f$.

One further step of simplification can be done by introducing the new coordinate $\tilde t(t) = \int_0^t d\lambda B(\lambda)$, which implies $\frac{d\tilde t}{dt} = B(t)$ and thus, for~the tangent bundle coordinates, $\dot{\tilde t} = \dot t B(t)$ and $\tilde s = w/\dot{\tilde t}$. In~these new coordinates \eqref{eq:Lnontrivial} becomes
\begin{align}\label{eq:Lnontrivial}
	L(\tilde t, \dot{\tilde t}, w) =  \dot {\tilde t}^2 f(\tilde s)\,.
\end{align}
eventually there is only one free function $f$, which needs to be determined by Finsler gravitational field equations. The~null structure of the above Finsler Lagrangian is determined by the zeros of $f$, and~surely $f$ has to be choosen in such a way that $L$ defines a Finsler spacetime according its definition in Section~\ref{sec:fins}.

To summarize our findings we have proven the following theorem:
\begin{Theorem}\label{Thm:1} If a Finsler spacetime Lagrangian $L$ is of Berwald type and admits cosmological, spatial~homogeneous and isotropic symmetry, then it falls into one of the following~classes:
\begin{enumerate}[leftmargin=*,labelsep=5mm]
	\item pseudo-Riemannian (quadratic in $\dot x$), in~which case it is, up~to $t$ coordinate redefinition, given by the FLRW metric, or~	\item nontrivially Finslerian, in~which case it is of the form~\eqref{eq:Lnontrivial}.
\end{enumerate}
\end{Theorem}

As an explicit example, one may consider $f = ( c - \tilde s^2) h(\tilde s)$ with $c$ being a constant and $h$ being a smooth and non-vanishing function. The~resulting Finsler Lagrangian then is
\begin{align}
	L = (c \dot {\tilde{t}}^2 - w^2 )h(\tilde s)\,,
\end{align}
which satisfies all Finsler spacetime criteria. 
Still, the~function $h$ can be chosen freely and must be determined from the gravitational field equation. The~latter is work in~progress.

%%%%%%%%%%%%%%%%%%%%%%%%%%%%%%%%%%%%%%%%%%%%%%%%%%%%%%%%%%%%%%%%%%%%%%%%%%%%%%%%%%%%%%%
\section{Discussion and~Conclusions}\label{sec:conc}
The cosmological principle assumes the existence of a symmetry group $G$ that acts transitively on spatial hypersurfaces of spacetime, and~which contains a local isotropy group $G_p\subset G$. In~the context of Finsler geometry, so far, it has been assumed that the generators of this symmetry group have the form \eqref{eq:iso} and \eqref{eq:trans}. So far this had not been derived from first principles. We closed this gap by showing that the dimension of $G$ must be $6$ and that the dimension of $G_p$ must be $3$. Thus the dimension of these groups is the same in the pseudo-Finsler and in the pseudo-Riemannian setting. Moreover, with~the help of an auxiliary metric on the spatial slices sharing the same symmetry group, which is guaranteed to exist, we could conclude that the symmetry generators indeed must have the assumed~form.

Among the variety of possible Finsler geometric extensions of pseudo-Riemannian geometry as geometry of spacetime, Berwald spacetimes represent a most conservative generalization. Our~discovery of the most general non-trivial cosmological, that is,\  spatially homogeneous and isotropic Berwald spacetimes reveals the class of geometries which extend the famous FLRW class of metrics into this realm. Most importantly, we found that cosmological Berwald geometries are parametrized by a free $0$-homogeneous function on the tangent bundle, which intertwines the position and direction dependence of the Finsler Lagrangian in a very specific way. The~resulting Finsler Lagrangian is
\begin{align}\label{eq:berwcosmo}
	L = \dot {\tilde t}^2 f(\tilde s)\,.
\end{align}

As the scale factor is determined by the Einstein equations on general relativity, the~free function must be determined by suitable Finsler generalisations of the Einstein equations. Most of the suggested generalizations in the literature simplify significantly for Berwald~geometries.

In particular, the~ansatz \eqref{eq:berwcosmo} is an important step in the program of the description of the evolution of the universe in terms of a gravitational field distribution sourced by a kinetic gas. We argued in Reference~\cite{Hohmann:2019sni} that the back reaction of a kinetic gas on the geometry of spacetime can be obtained directly from the $1$-particle distribution function (1PDF) of the gas, when one employs Finsler geometry instead of pseudo-Riemannian geometry. The~explicit form of the 1PDF will then determine the free function $f$, a~derivation which is currently work in~progress.

%%%%%%%%%%%%%%%%%%%%%%%%%%%%%%%%%%%%%%%%%%
 \vspace{6pt}
\authorcontributions{The authors have all contributed substantially to the derivation of the presented results as well as analysis, drafting, review, and~finalization of the manuscript. All authors have read and agreed to the published version of the~manuscript.}

%%%%%%%%%%%%%%%%%%%%%%%%%%%%%%%%%%%%%%%%%%
\funding{C.P. and M.H. were supported by the Estonian Ministry for Education and Science through the Personal Research Funding Grants PSG489 (C.P.) and PRG356 (M.H.), as~well as the European Regional Development Fund through the Center of Excellence TK133 ``The Dark Side of the Universe''.}

%%%%%%%%%%%%%%%%%%%%%%%%%%%%%%%%%%%%%%%%%%
\acknowledgments{The authors would like to acknowledge networking support by the COST Actions QGMM (CA18108) and CANTATA (CA15117), supported by COST (European Cooperation in Science and Technology).}

%%%%%%%%%%%%%%%%%%%%%%%%%%%%%%%%%%%%%%%%%%
\conflictsofinterest{The authors declare no conflict of~interest.}

%%%%%%%%%%%%%%%%%%%%%%%%%%%%%%%%%%%%%%%%%%
%% optional
%\appendixtitles{no} % Leave argument "no" if all appendix headings stay EMPTY (then no dot is printed after "Appendix A"). If the appendix sections contain a heading then change the argument to "yes".
%\appendix

%\section{An Appx}\label{app:1}

%%%%%%%%%%%%%%%%%%%%%%%%%%%%%%%%%%%%%%%%%%
\reftitle{References}


%\externalbibliography{yes}
%\bibliography{CBFST}

%%%%%%%%%%%%%%%%%%%%%%%%%%%%%%%%%%%%%%%%%%
\end{document}